\documentclass{fic-l}

\theoremstyle{definition}

\theoremstyle{remark}

\numberwithin{equation}{section}

\usepackage{graphicx}
\usepackage{amssymb}

\begin{document}

\def\star{{\displaystyle *}}
\def\be{\begin{equation}}
\def\ee{\end{equation}}

\title{How to model quantum plasmas}

\author{Giovanni Manfredi}
\address{Laboratoire de Physique des Milieux Ionis{\'e}s et Applications\\ CNRS and Universit{\'e}
Henri Poincar{\'e}, BP 239\\ F-54506 Vandoeuvre-les-Nancy\\ France}


\begin{abstract}
Traditional plasma physics has mainly focused on regimes
characterized by high temperatures and low densities, for which
quantum-mechanical effects have virtually no impact. However,
recent technological advances (particularly on miniaturized
semiconductor devices and nanoscale objects) have made it possible
to envisage practical applications of plasma physics where the
quantum nature of the particles plays a crucial role. Here, I
shall review different approaches to the modeling of quantum
effects in electrostatic collisionless plasmas. The full kinetic
model is provided by the Wigner equation, which is the quantum
analog of the Vlasov equation. The Wigner formalism is
particularly attractive, as it recasts quantum mechanics in the
familiar classical phase space, although this comes at the cost of
dealing with negative distribution functions. Equivalently, the
Wigner model can be expressed in terms of $N$ one-particle
Schr{\"o}dinger equations, coupled by Poisson's equation: this is
the Hartree formalism, which is related to the `multi-stream'
approach of classical plasma physics. In order to reduce the
complexity of the above approaches, it is possible to develop a
quantum fluid model by taking velocity-space moments of the Wigner
equation. Finally, certain regimes at large excitation energies
can be described by semiclassical kinetic models (Vlasov-Poisson),
provided that the initial ground-state equilibrium is treated
quantum-mechanically. The above models are validated and compared
both in the linear and nonlinear regimes.

\end{abstract}

\maketitle

\section{Introduction}
Plasma physics deals with the $N$-body dynamics of a system of
charged particles interacting via electromagnetic forces. The
study of plasmas arose in the early twentieth century when
scientists got interested in the physics of gas discharges. After
World War II, plasmas became the object of intensive experimental
and theoretical research, mainly because of the potential
applications of nuclear fusion, both military (hydrogen bomb) and
peaceful (energy production through controlled thermonuclear
fusion). In parallel, plasma physics was also developed by
astrophysicists and geophysicists, which is not surprising, as it
is thought that about $90 \%$ of all matter in the visible
universe exists in the form of a plasma. More precisely, plasmas
are observed in the Sun surface, the earth's magnetosphere, and
the interplanetary and interstellar media.

Both fusion and space plasmas are characterized by regimes of high
temperature and low density, for which quantum effects are totally
negligible \footnote{Quantum effects do play an important role,
for example, in determining the fusion cross sections. But this is
a nuclear physics rather than a plasma physics issue. What we mean
is that the {\it dynamics} and {\it thermodynamics} of fusion and
space plasmas are completely unaffected by quantum corrections.}.
However, physical systems where both plasma and quantum effects
coexist do occur in nature, the most obvious example being the
electron gas in an ordinary metal. In metals, valence electrons
are not attached to a particular nucleus, but rather behave as
free particles, which is why metals make good electric conductors.
Although some level of understanding of metallic properties is
achieved by considering noninteracting electrons, a more accurate
description can be obtained by treating the electron population as
a plasma, globally neutralized by the lattice ions. At room
temperature and standard metallic densities, quantum effects can
no longer be ignored, so that the electron gas constitutes a true
{\it quantum plasma}.

For ordinary metals, however, the properties of the electron
population (band structure, thermodynamic properties) are mainly
determined by the presence of a regular ion lattice, typical
plasma effects being only a higher order correction. In recent
years, though, there has been tremendous progress in the
manipulation of metallic nanostructures (metal clusters,
nanoparticles, thin metal films) constituted of a small number of
atoms (typically $10-10^5$)
\cite{Calvayrac,Brorson,Suarez,Bigot,Domps2}. For such objects, no
underlying ionic lattice exists, so that the dynamics of the
electron population is principally governed by plasma effects, at
least for large enough systems. Further, the development of
ultrafast (femtosecond and, more recently, attosecond) laser
sources makes it possible to probe the electron dynamics in
metallic nanostructures on the typical time scale of plasma
phenomena, which is indeed of the order of the femtosecond.
Metallic nanostructures thus constitute an ideal arena to study
the dynamical properties of quantum plasmas.

Another possible application of quantum plasmas arises from
semiconductor physics
\cite{Kluksdahl,Markowich,Yala,Luscombe,Arnold}. Even though the
electron density in semiconductors is much lower than in metals,
the great degree of miniaturization of today's electronic
components is such that the de Broglie wavelength of the charge
carriers can be comparable to the spatial variation of the doping
profiles. Hence, typical quantum mechanical effects, such as
tunneling, are expected to play a central role in the behavior of
electronic components to be constructed in the next years.

Finally, quantum plasmas also occur in some astrophysical objects
under extreme conditions of temperature and density, such as white
dwarf stars, where the density is some ten order of magnitudes
larger than that of ordinary solids \cite{Balberg}. Because of
such large densities, a white dwarf can be as hot as a fusion
plasma ($10^8$ K), but still behave quantum-mechanically.

A classical system of charged particles qualifies as a plasma if
it is {\it quasineutral} and if {\it collective effects} play a
significant role in the dynamics \cite{Chen}. `Quasineutrality'
means that charge separation can only exist on a short distance,
which, for classical plasmas, is given by the Debye length. On
distances larger than the Debye length, the plasma is basically
neutral, except for small fluctuations.
By `collective effects' we mean particle motions that depend not
only on local conditions, but also -- indeed principally -- on the
positions and velocities of all other particles in the plasma. (In
solid state and nuclear physics, collective effects are usually
called `mean-field' effects, because they arise from the average
field created by all the particles). Such collective behavior is
possible because of the long-range nature of electromagnetic
forces. In contrast, for neutral gases, the dominant interaction
mechanism is provided by short-range molecular forces of the
Lennard-Jones type: individual molecules thus move undisturbed in
the gas until they make a collision with another molecule (which
occurs when the interaction potentials overlap).
It should be added that real plasmas are often only partially
ionized, so that a fraction of neutral molecules is also present.
In order to have true plasma behavior, we must therefore require
that the collision rates of electrons and ions with the neutral
molecules be relatively low compared to typical collective
phenomena. In the present work, we shall avoid this issue
altogether by restricting our analysis to the simpler case of
fully ionized gases.

When quantum effects start playing a role, the above picture gets
more complicated, as an additional length scale is introduced,
namely the de Broglie wavelength of the charged particles,
$\lambda_B = \hbar/mv$. The de Broglie wavelength roughly
represents the spatial extension of the particle wave function --
the larger it is, the more important quantum effects are. From the
definition of $\lambda_B$, it is clear that quantum behavior will
be reached much more easily for the electrons than for the ions,
due to the large mass difference. Indeed, in all practical
situations, even the most extreme, the ion dynamics is always
classical, and only the electrons need to be treated
quantum-mechanically. In the present paper, we shall always refer
implicitly to electrons when discussing quantum effects. In
addition, only electrostatic (Coulomb) interactions will be
considered. Magnetic fields do introduce novel and interesting
effects, but the fundamental properties of quantum plasmas are
already present in the purely electrostatic scenario.

In the rest of this paper, we shall first obtain a number of
qualitative results by using simple arguments from dimensional
analysis. This will be useful to extract the relevant
dimensionless parameters that determine the various physical
regimes (classical/quantum, collisionless/collisional).
Subsequently, we shall derive and illustrate several mathematical
models that are appropriate to describe the dynamics of a quantum
plasma in the collisionless regime.

\section{Physical regimes for classical and quantum plasmas}
In this section, we shall derive a number of parameters that
represent the typical length, time, and velocity scales in a
classical or quantum plasma. These can be obtained using
elementary considerations based on dimensional analysis. Of
course, more detailed studies would be necessary to understand how
such parameters actually intervene in real physical phenomena (for
instance, whether a certain time scale represents a typical
oscillation frequency, or rather a damping rate). Here, we shall
derive the algebraic expression for these quantities and simply
state, without proof, what they represent physically.

In addition, it will also be important to establish certain {\it
dimensionless} parameters. Dimensionless parameters allow us to
discriminate between different physical regimes, characterized by
situations where one effect dominates over another. In particular,
we shall look for parameters that define whether a plasma is
classical or quantum, and whether it is dominated by individual
effects ({\it collisional}) or collective effects ({\it
collisionless}).

\subsection{Classical plasmas}
We consider a plasma of number density $n$, composed of particles
(typically, electrons) with mass $m$ and electric charge $e$,
interacting via Coulomb forces (hence the electric permittivity
$\epsilon_0$). With these four parameters, we are able to
construct a quantity that has the dimensions of an inverse time,
i.e. a frequency:
\be \omega_p = \left( \frac{e^2n}{m\varepsilon_0}\right)^{1/2}.
\label{omegap} \ee

The latter quantity is known as the {\it plasma frequency} and it
represents the typical oscillation frequency for electrons
immersed in a neutralizing background of positive ions, which is
supposed to be motionless because of the large ion mass. The
oscillations arise from the fact that, when a portion of the
plasma is depleted of some electrons (thus creating a net positive
charge), the resulting Coulomb force tends to pull back the
electrons towards the excess positive charge. Due to their
inertia, the electrons will not simply replenish the positive
region, but travel further away thus re-creating an excess
positive charge. In the absence of collisions, this effect gives
rise to undamped electron oscillations at the plasma frequency.

Note that the plasma frequency is independent on the temperature.
If we do introduce a finite temperature $T$, then we can construct
a typical velocity:
\be
v_T = \left( \frac{k_B T}{m}\right)^{1/2},
\label{vt}
\ee
where $k_B$ is Boltzmann's constant.
This is the {\it thermal velocity}, which represents, just like in
ordinary gases, the typical speed due to random thermal motion.

By combining the above two quantities, one can define a typical
length scale, the {\it Debye length}:
\be \lambda_D = \frac{v_T}{\omega_p} = \left(
\frac{\varepsilon_{0} k_B T}{n e^2}\right)^{1/2}.
\label{lambdad}
\ee
The Debye length describes the important phenomenon of
electrostatic screening: if an excess positive charge is
introduced in the plasma, it will be rapidly surrounded by a cloud
of electrons (which are more mobile and thus react quickly). As a
result, the positive charge will be partially screened and will be
virtually `invisible' to other particles situated at a large
enough distance. Quantitatively, this amounts to saying that the
electrostatic potential generated by an excess charge does not
fall, like in vacuum, as $1/r$, but rather obeys a Yukawa-like
potential $\exp(-r/\lambda_D)/r$, which of course decays much more
quickly and on a distance of the order of the Debye length. The
Debye screening is at the origin of one of the most crucial of all
plasma properties, namely {\it quasineutrality}: charge separation
in a plasma can exist only on scales smaller than $\lambda_D$, but
it is screened out at larger sales.

Let us now try to construct a dimensionless parameter using the
above quantities: $m,~e,~\varepsilon_0,~n$, and $T$. It is easily
seen that only one such parameter exists, and it reads as
\be g_C = \frac{e^2 n^{1/3}}{\varepsilon_0 k_B T}~. \label{gc1}
\ee

This is known as the (classical) {\it graininess parameter} or
{\it coupling parameter}. It is illuminating to show that $g_C$
can be written as the ratio of the interaction (electric) energy
$E_{\rm int}$ to the average kinetic energy $E_{\rm kin}$. Indeed,
for particles situated at typical interparticle distance
$d=n^{-1/3}$, one has $E_{\rm int}= e^2 /(\varepsilon_0 d)$ and
$E_{\rm kin} = k_B T$, which immediately yields the expression
(\ref{gc1}).

The expression $g_C = E_{\rm int}/E_{\rm kin}$ allows us to guess
the physical relevance of the coupling parameter. When $g_C$ is
small, the plasma is dominated by thermal effects, whereas
two-body Coulomb interactions (i.e. binary collisions) remain
weak. In this regime, the main field acting on the charged
particles is the nonlocal mean field, which is responsible for
typical collective effects. This is known as the {\it
collisionless} regime. On the contrary, when $g_C \simeq 1$ or
larger, binary collisions cannot be neglected and the plasma is
said to be {\it collisional} or {\it strongly coupled}. We also
note that, following (\ref{gc1}), classical plasmas are
collisionless at high temperatures and low densities.

Alternatively, $g_C$ can be written as the inverse of the number
of particles contained in a volume of linear dimension
$\lambda_D$, raised to a certain power:
\be g_C = \left( \frac{1}{n\lambda_D^3}\right)^{2/3}.
\label{gc2}\ee
This shows that a plasma is collisionless when the Debye screening
is effective, i.e. when a large number of electrons are available
in a Debye volume.

\subsection{Quantum plasmas}\label{sec:quantum plasmas}
Quantum effects can be measured by the thermal de Broglie
wavelength of the particles composing the plasma
\be \lambda_B = \frac{\hbar}{m v_T} \label{lambdab}~, \ee
which roughly represents the spatial extension of a particle's
wave function due to quantum uncertainty. For classical regimes,
the de Broglie wavelength is so small that particles can be
considered as pointlike (except, as mentioned in the Introduction,
when computing collision cross-sections), therefore there is no
overlapping of the wave functions and no quantum interference. On
this basis, it is reasonable to postulate that quantum effects
start playing a significant role when the de Broglie wavelength is
similar to or larger than the average interparticle distance
$n^{-1/3}$, i.e. when
\be n\lambda_B^3 \geq 1~. \label{chi1} \ee

On the other hand, it is well known from the statistical mechanics
of ordinary gases \cite{Landau} that quantum effects become
important when the temperature is lower than the so-called Fermi
temperature $T_F$, defined as
\be
k_B T_F \equiv E_F = \frac{\hbar^2}{2m}~(3\pi^2)^{2/3}~ n^{2/3}~,
\label{tf}
\ee
where we have also defined the Fermi energy $E_F$. When $T$
approaches $T_F$, the relevant statistical distribution changes
from Maxwell-Boltzmann to Fermi-Dirac. Now, it is easy to see that
the ratio $\chi \equiv T/T_F$ is simply related to the
dimensionless parameter $n\lambda_B^3$ discussed above:
\be \chi \equiv \frac{T_F}{T} = \frac{1}{2}~(3\pi^2)^{2/3} ~(n
\lambda_B^3)^{2/3}. \label{chi2} \ee
Thus, quantum effects become important when $\chi \ge 1$.

We now want to establish the typical space, time, and velocity
scales for a quantum plasma, as well as the relevant dimensionless
parameters. First of all, we stress that simple expressions can be
found only in the limiting cases $T \gg T_F$ (corresponding to the
classical case treated previously) and $T \ll T_F$, which is the
`deeply quantum' (fully degenerate) regime that we are going to
analyze. Of course, there will be a smooth transition between the
two regimes, but this cannot be treated using straightforward
dimensional arguments.

Concerning the typical time scale for collective phenomena, this
is still given by the inverse of the plasma frequency
(\ref{omegap}), even in the quantum regime. However, the thermal
speed becomes meaningless in the very low temperature limit, and
should be replaced by the typical velocity characterizing a
Fermi-Dirac distribution. This is the Fermi velocity:
\be v_F = \left(\frac{2 E_F}{m}\right)^{1/2} = \frac{\hbar}{m}~(3
\pi^2~ n)^{1/3}~. \label{vf} \ee

With the plasma frequency and the Fermi velocity, we can define a
typical length scale
\be \lambda_F = \frac{v_F}{\omega_p}~, \label{lambdaf}\ee
which is the quantum analog of the Debye length. Just like the
Debye length, $\lambda_F$ describes the scale length of
electrostatic screening in a quantum plasma.

The quantum coupling parameter can be defined as the ratio of the
interaction energy $E_{\rm int}$ to the average kinetic energy
$E_{\rm kin}$. The interaction energy is the same as in the
classical case, whereas the kinetic energy is now given by the
Fermi energy $E_{\rm kin} = E_F$. With these assumptions, one can
write the quantum coupling parameter as
\be g_Q \equiv \frac{E_{\rm int}}{E_F} =  \frac{2}{(3\pi^2)^{2/3}}~
\frac{e^2 m }{\hbar^2 \varepsilon_0 ~n^{1/3}} \sim \left(
\frac{1}{n \lambda_F^3}\right)^{2/3} \sim \left(\frac{\hbar
\omega_p}{E_F}\right)^2 ~, \label{gq}\ee
where we have left out proportionality constants for sake of
clarity. The third expression of $g_Q$ in (\ref{gq}) is completely
analogous to the classical one when one substitutes $\lambda_F \to
\lambda_D$. The last expression is more interesting, as it has no
classical counterpart: it describes the coupling parameter as the
ratio of the `plasmon energy' $\hbar \omega_p$ (energy of an
elementary excitation associated to an electron plasma wave) to
the Fermi energy \footnote{$g_Q$ is directly proportional to the
parameter $r_s/a_0$ (where $a_0$ is Bohr's radius), commonly used
in solid state physics \cite{Ashcroft}.}.

The quantum collisionless regime (where collective, mean-field
effects dominate) is again defined as the regime where the quantum
coupling parameter is small. From (\ref{gq}), it appears that a
quantum plasma is `more collective' at larger densities, in
contrast to a classical plasma [see (\ref{gc1})]. This may seem
surprising, but can be easily understood by invoking Pauli's
exclusion principle, according to which two fermions cannot occupy
the same quantum state. In a fully degenerate fermion gas, all
low-energy states are occupied: if we add one more particle to the
gas, it will necessarily be in a high-energy state. Therefore, by
increasing the gas density, we automatically increase its average
kinetic energy, which, in virtue of (\ref{gq}), reduces the value
of $g_Q$.

\subsection{Plasma regimes}
We have so far defined three dimensionless parameters that
determine whether the plasma is classical or quantum, and, in
either case, whether it is collisional or collisionless:

\begin{enumerate}
\item $\chi = T_F/T$ : classical/quantum
\item $g_C$ : collisional/collisionless (classical regime)
\item $g_Q$ : collisional/collisionless (quantum regime)
\end{enumerate}

\begin{figure}[tb]
\begin{center}
\includegraphics*[width=8cm]{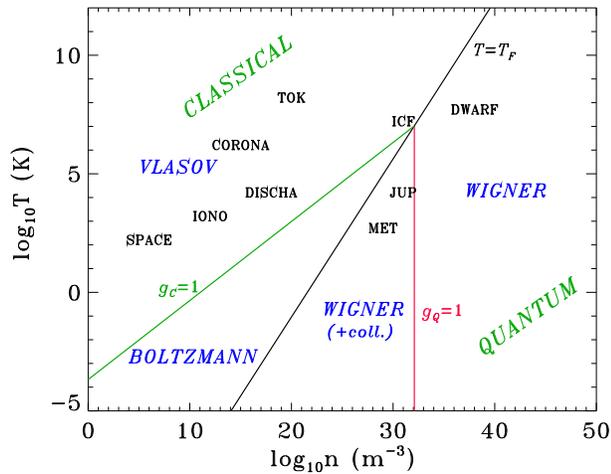}
\end{center}
\vspace{0pc} \caption{Plasma diagram in the $\log T - \log n$
plane. IONO: ionospheric plasma; SPACE: interstellar space;
CORONA: solar corona; DISCHA: typical electric discharge; TOK:
tokamak experiment (magnetic confinement fusion); ICF: inertial
confinement fusion; MET: metals and metal clusters; JUP: Jupiter's
core; DWARF: white dwarf star.} \label{fig:logt}
\end{figure}

These parameters are functions of the temperature and density. In
Fig. \ref{fig:logt}, we plot on a $\log T - \log n$ diagram the
straight lines corresponding to $\chi = g_C = g_Q = 1$, which
delimitate the various plasma regimes \cite{Bonitz}.

The $\log T - \log n$ plane is divided into four regions, two of
which are classical (above the $T=T_F$ line) and two quantum. Each
quantum/classical region is then divided in two
collisional/collisionless subregions, identified by the kinetic
equations that are relevant to each regime. As we shall see in the
forthcoming sections, the Vlasov and Wigner equations are the
appropriate models respectively for collisionless classical and
collisionless quantum plasmas. `Boltzmann' is used as a generic
term for collisional kinetic equations in the classical regime.
Collisional effects in the quantum regime are much harder to deal
with, and no uncontroversial kinetic model exists in that regime,
which is identified by `Wigner (+ coll.)' on the figure.

We point out that all previous considerations have implicitly
assumed thermal equilibrium. Out-of-equilibrium regimes should be
treated much more carefully and the above results may not be
entirely correct. For example, if an electron beam is injected
into a plasma, the beam velocity will have to be taken into
account when determining the de Broglie wavelength, which will
therefore be smaller. For this reason, systems that are far from
equilibrium can sometimes be treated with a semiclassical model,
even though the corresponding equilibrium may still be fully
quantum (see Sec. \ref{sec:Vlasov}).

Several points corresponding to natural and laboratory plasmas
have also been plotted in Fig. 1. We note that space and magnetic
fusion plasmas fall in the classical collisionless region, whereas
inertial confinement fusion plasmas may display quantum and/or
strong coupling effects. Extremely dense astrophysical objects
such as white dwarf stars are definitely quantum and
collisionless, even though they are as hot as fusion or solar
plasmas.

\section{Electrons in metals and metallic nanostructures: Pauli blocking}
\label{Pauli} 
\begin{figure}[tb]
\begin{center}
\includegraphics*[width=6cm]{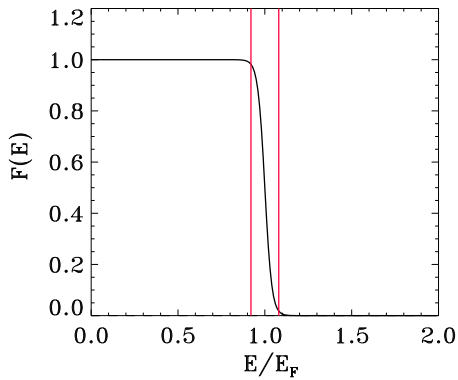}
\end{center}
\vspace{0pc} \caption{Fermi-Dirac energy distribution for a case
with $T/T_F = 0.1$. The vertical lines define a shell of thickness
$k_B T$ around the Fermi surface.} \label{fig:fd}
\end{figure}

The typical quantum coupling parameter for ordinary
metals is larger than unity, so that, in principle,
electron-electron collisions are as important as collective
effects. If that were really the case, one should abandon
collisionless models altogether and resort to the full $N$-body
problem. This is hardly a feasible task. Fortunately, however, the
effect known as {\it Pauli blocking} reduces the collision rate
quite dramatically in most cases of interest. This occurs when the
electron distribution is close to the Fermi-Dirac equilibrium at
relatively low temperatures. The fundamental point is that, when
all lower levels are occupied, the exclusion principle forbids a
vast number of transitions that would otherwise be possible
\cite{Ashcroft}. In particular, at strictly zero temperature, all
electrons have energies below $E_F$, and no transition is
possible, simply because there are no available states for the
electrons to occupy. At moderate temperatures, only electrons
within a shell of thickness $k_B T$ about the Fermi surface (i.e.
the region where $E=E_F$) can undergo collisions (this shell is
delimited by the two vertical lines in Fig. \ref{fig:fd}). For
such electrons, the e-e collision rate (inverse of the lifetime
$\tau_{ee}$) is proportional to $k_B T/\hbar$ (this is a form of
the uncertainty principle, energy $\times$ time = const.). The
{\it average} collision rate is obtained by multiplying the
previous expression by the fraction of electrons present in the
shell of thickness $T$ about the Fermi surface, which is $\sim
T/T_F$. One obtains
\be \label{tauee} \nu_{ee} \sim ~\frac{k_B T^2}{\hbar ~T_F}~. \ee
In normalized units, this expression reads as \be
\frac{\nu_{ee}}{\omega_p} = \frac{E_F}{\hbar
\omega_p}~\left(\frac{T}{ T_F}\right)^2 =
\frac{1}{g_Q^{1/2}}~\left(\frac{T}{T_F}\right)^2 \label{nuee} \ee
Thus, $\nu_{ee} < \omega_p$ in the region where $T<T_F$ and
$g_Q>1$, which is the relevant one for metallic electrons (Fig.
\ref{fig:logt}). Restoring dimensional units, we find that, at
room temperature, $\tau_{ee} \simeq 10^{-10}~{\rm s}$, which is
much larger than the typical collisionless time scale $\tau_p = 2
\pi \omega_p^{-1} \simeq 10^{-15}~{\rm s} = 1~{\rm fs}$. In
addition, the typical time scale for electron-lattice collisions
$\tau_{ei} \simeq 10^{-14}~{\rm s} = 10~{\rm fs}$ is also larger
than $\tau_p$. Therefore, it appears that a collisionless regime
is indeed relevant on a time scale of the order of the
femtosecond.

A word of caution is in order, however, not to overestimate the
effect of Pauli blocking. The above considerations are valid at
thermodynamic equilibrium, whereas many more transitions are
allowed for out-of-equilibrium electrons. Therefore, for strongly
excited systems where many nonequilibrium electrons are present,
the collision frequency may be larger than the simple estimate
given in (\ref{nuee}).

Some typical parameters for metallic electrons (gold) at room
temperature are summarized in Table 1. We note that $\tau_{p}$,
the typical collisionless time scale, is of the order of the
femtosecond. With the recent development of ultrafast laser
sources with femtosecond period, it is therefore possible to probe
the mean-field properties of metallic nanostructures. For example,
the electron dynamics in thin gold films excited with femtosecond
lasers was studied experimentally in several
works~\cite{Brorson,Suarez}. We also point out that $\lambda_F$,
the typical collisionless length scale, is of the order of the \AA
ngstrom, which is comparable to the typical atomic size.

\begin{table}[ht]
\caption{Typical parameters for electrons in gold at room
temperature}
\begin{center}
\begin{tabular}{|c|c|}
\hline
  $n$ &  $5.9 \times 10^{28} ~{\rm m}^{-3}$  \\ \hline
  $T$ & $300$ K \\ \hline
  $\omega_{p}$  & $1.37 \times 10^{16} ~{\rm s^{-1}}$  \\ \hline
  $\tau_{p}$ &  $0.46 ~{\rm fs}$ \\ \hline
  $\tau_{ie}$ &  $30 ~{\rm fs}$ \\ \hline
  $T_F$ & $6.4 \times 10^{4} ~{\rm K}$  \\ \hline
  $E_F$ & $5.53 ~{\rm eV}$  \\ \hline
  $v_F$ & $1.4 \times 10^{6} ~{\rm ms^{-1}}$  \\ \hline
  $\lambda_F$ & $1 \times 10^{-10} ~{\rm m}$ \\ \hline
  $g_Q$ & 12.7 \\ \hline
  $r_s/a_0$ & 3 \\ \hline
\end{tabular}
\end{center}
\end{table}

\section{Models}\label{sec:Models}
The most fundamental model for the quantum $N$-body problem is the
Schr\"{o}din- ger equation for the $N$-particle wave function
$\psi(x_1,x_2,\dots,x_N,t)$. Obviously, this is an unrealistic
task, both for analytical calculations and numerical simulations.
A drastic, but useful and to some extent plausible, simplification
can be achieved by neglecting two-body (and higher order)
correlations. This amounts to assume that the $N$-body wave
function can be factored into the product of $N$ one-body
functions: \be \psi(x_1,x_2,\dots,x_N,t) =
\psi_1(x_1,t)~\psi_2(x_2,t)\dots \psi_N(x_N,t). \label{nbody} \ee
For fermions, a weak form of the exclusion principle is satisfied
if none of the wave functions on the right-hand side of
(\ref{nbody}) are identical \footnote{A stronger version of the
exclusion principle requires that $\psi(x_1,x_2,\dots,x_N,t)$ is
antisymmetric, i.e. that it changes sign when two of its arguments
are interchanged. This can be achieved by taking, instead of the
single product of $N$ wave functions as in (\ref{nbody}), a linear
combinations of all products obtained by permutations of the
arguments, with weights $\pm 1$ (Slater determinant)
\cite{Ashcroft}. This is at the basis of Fock's generalization of
the Hartree model, as pointed out in Sec. \ref{sec:hartree}.}.

The assumption of weak correlation between the particles is
satisfied, as we have seen, when the quantum coupling parameter
$g_Q$ is small. The set of $N$ one-body wave functions is known as
a quantum mixture (or quantum mixed state) and is usually
represented by a density matrix
\be \rho(x,y,t) =
\sum_{\alpha=1}^{N} p_\alpha  \psi_\alpha(x,t)
~\psi_\alpha^{\displaystyle*}(y,t), \ee
where, for clarity, we have assumed the same normalization $\int
\vline \psi_\alpha \vline^{2} dx =1$ for all wave functions and then introduced the occupation
probabilities $p_\alpha$.

Both the Wigner and Hartree models described below are completely
equivalent to models based on the density matrix formalism (Von
Neumann equation).

\subsection{Wigner-Poisson}
The Wigner representation \cite{Wigner} is a useful tool to
express quantum mechanics in a phase space formalism (for reviews
see \cite{Moyal,Tatarskii,Hillery}). As detailed above, a generic
quantum mixed state can be described by $N$ single-particle wave
functions $\psi_\alpha(x,t)$ each characterized by a probability
$p_\alpha$ satisfying $\sum_{\alpha=1}^N  p_\alpha =1$. The Wigner
function is a function of the phase space variables $(x, v)$ and
time, which, in terms of the single-particle wave functions, reads
as

\be f(x,v,t) = \sum_{\alpha=1}^{N} \frac{m}{2\pi\hbar}~ p_\alpha
\int_{-\infty}^{+\infty} \psi_\alpha^{\displaystyle*} \left(x +
\frac{\lambda}{2},t\right) \psi_\alpha\left(x -
\frac{\lambda}{2},t\right) e^{i m v\lambda/\hbar}~d\lambda
\label{wigfunc} \ee
(we restrict our discussion to one-dimensional cases, but all
results can easily be generalized to three dimensions). It must be
stressed that the Wigner function, although it possesses many
useful properties, is not a true probability density, as it can
take negative values. However, it can be used to compute averages
just like in classical statistical mechanics. For example, the
expectation value of a generic quantity $A(x,v)$ is defined as:
\be \langle A \rangle = \frac{\int \int f(x,v) A(x,v) dx dv}{\int \int
f(x,v) dx dv}, \label{average}\ee
and yields the correct quantum-mechanical value \footnote{For
variables whose corresponding quantum operators do not commute
(such as $\hat{x}\hat{v}$), (\ref{average}) must be supplemented
by an ordering rule, known as Weyl's rule \cite{Hillery}.}. In
addition, the Wigner function reproduces the correct
quantum-mechanical marginal distributions, such as the spatial
density: \be n(x,t) = \int_{-\infty}^{+\infty} f(x,v,t) ~dv =
\sum_{\alpha=1}^N p_\alpha \mid \psi_\alpha \mid^2. \ee

We also point out that, of course, not all functions of the phase
space variables are genuine Wigner functions, as they cannot
necessarily be written in the form (\ref{wigfunc}). In general,
although it is trivial to find the Wigner function given the $N$
wave functions that define the quantum mixture, the inverse
operation is not generally feasible. Indeed, there are no simple
rules to establish whether a given function of $x$ and $v$ is a
genuine Wigner function. For a more detailed discussion on this
issue, and some practical recipes to construct genuine Wigner
functions, see \cite{Manfredi}.

The Wigner function obeys the following evolution equation:
\[
\frac{\partial{f}}{\partial{t}} + v\frac{\partial{f}}{\partial{x}}
~+
\]
\be \frac{em}{2i\pi\hbar^2}
\int\int{d\lambda}~{dv'}e^{im(v-v')\lambda/\hbar} \label{wignereq}
\left[\phi\left(x+\frac{\lambda}{2}\right)-
\phi\left(x-\frac{\lambda}{2}\right)\right]f(x,v',t) = 0~, \ee
where $\phi(x,t)$ is the self-consistent electrostatic potential.
Developing the integral term in (\ref{wignereq}) up to order
$O(\hbar^2)$ we obtain \be \frac{\partial{f}}{\partial{t}} +
v\frac{\partial{f}}{\partial{x}}- \frac{e}{m} \frac{\partial
\phi}{\partial{x}} \frac{\partial{f}}{\partial{v}} =
\frac{e\hbar^2}{24 m^3} \frac{\partial^3 \phi}{\partial{x^3}}
\frac{\partial^3{f}}{\partial{v^3}}+ O(\hbar^4) \ee
The Vlasov equation is thus recovered in the formal semiclassical
limit $\hbar \to 0$. We stress, however, that rigorous asymptotic
results are much harder to obtain and generally involve weak
convergence.

The Wigner equation must be coupled to the Poisson's equation for
the electric potential
\be \label{poisson} \frac{\partial^2 \phi}{\partial x^2} =
\frac{e}{\varepsilon_0}\left(\int f dv - n_0\right)~, \ee
where we have assumed that the ions form a motionless neutralizing
background with uniform density $n_0$ (this is known as the
`jellium' model in solid state physics).

The resulting Wigner-Poisson (WP) system has been extensively used
in the study of quantum transport
\cite{Kluksdahl,Markowich,Drummond}. Exact analytical results can
be obtained by linearizing (\ref{wignereq}) and (\ref{poisson})
around a spatially homogeneous equilibrium given by $f_0(v)$. By
expressing the fluctuating quantities as a sum of plane waves
$\exp(i k x -i \omega t)$ with frequency $\omega$ and wave number
$k$, the dispersion relation can be written in the form
$\varepsilon(k,\omega)=0$, where the dielectric constant
$\varepsilon$ reads, for the WP system,
\be \varepsilon_{\rm WP}(\omega, k) = 1 + \frac{m \omega_p^2}{n_0
k} \int \frac{f_0(v+\hbar k/2m) - f_0(v-\hbar k/2m)}{\hbar k
(\omega-k v)}~dv ~. \label{dispwp1}\ee
With an appropriate change of integration variable,
(\ref{dispwp1}) can be written as
\begin{equation}
\label{dispwp2} \varepsilon_{\rm WP}(\omega, k) = 1 -
\frac{\omega_{p}^2}{n_{0}}\int\frac{f_{0}(v)} {(\omega -
kv)^{2} - \hbar^{2}k^{4}/4m^{2}}~dv~.
\end{equation}
From (\ref{dispwp1}), one can recover the Vlasov-Poisson (VP)
dispersion relation by taking the semiclassical limit $\hbar\to 0$
\be \varepsilon_{\rm VP}(\omega, k) = 1 + \frac{\omega_p^2}{n_0 k}
\int \frac{\partial f_0/\partial v}{\omega-k v}~dv ~.
\label{dispvp}\ee

The integration in (\ref{dispwp1}) and (\ref{dispvp}) should be
performed along the Landau contour in the complex (Re$~v$, Im$~v$)
plane, so that the singularity at $v=\omega/k$ is always left
above the contour \cite{Chen}. This prescription allows one to
obtain the correct imaginary part of $\varepsilon(k,\omega)$,
which is at the basis of the phenomenon of Landau damping.

Just like in the classical case (Vlasov-Poisson), the dispersion
relation (\ref{dispwp1}) can also support unstable solutions, i.e.
solutions with a positive imaginary part of the frequency. These
solutions grow exponentially until some nonlinear effect kicks in
and leads to saturation of the instability. The stability property
of the dispersion relation of the Wigner-Poisson system have been
extensively studied in \cite{Haas2,Bonitz3}.

\subsection{Hartree}
\label{sec:hartree} A completely equivalent approach to the WP
system is obtained by making direct use of the $N$ wave functions
$\psi_\alpha(x,t)$. These obey $N$ independent Schr\"{o}dinger
equations, coupled through Poisson's equation

\begin{eqnarray}
i\hbar\frac{\partial\psi_\alpha}{\partial\,t} &=& -
\frac{\hbar^2}{2m}\frac{\partial^{2}\psi_\alpha}{\partial x^2}
- e\phi\psi_\alpha~, ~~~~~ \alpha = 1 \dots N \label{sp1}\\
\frac{\partial^2 \phi}{\partial x^2} &=&
\frac{e}{\varepsilon_0}\left(\sum_{\alpha=1}^{N} p_\alpha
|\psi_{\alpha}|^2 - n_{0}\right) \label{sp2}~.
\end{eqnarray}
This type of model was originally derived by Hartree in the
context of atomic physics, with the aim of studying the
self-consistent effect of atomic electrons on the Coulomb
potential of the nucleus. Subsequently, Fock introduced a
correction that accounts for the parity of the $N$-particle wave
function for an ensemble of fermions (Hartree-Fock model), but
this development will not be considered in this paper \footnote{A
very similar model to the Hartree equations
(\ref{sp1})--(\ref{sp2}) is known as TDDFT (time-dependent density
functional theory) \cite{Gross}. Its linearized version goes under
the name of RPA (random-phase approximation) \cite{Pines}.}.

Instead, it is useful to think of the above
Schr\"{o}dinger-Poisson equations (\ref{sp1})--(\ref{sp2}) as the
quantum-mechanical analog of Dawson's {\it multistream} model
\cite{Dawson}. Dawson supposed that the classical distribution
function can be represented as a sum of $N$ `streams', each
characterized by a probability $p_\alpha$, a density $n_\alpha$,
and a velocity $u_\alpha$:
\be f(x,v,t) = \sum_{\alpha=1}^{N} p_\alpha~
n_{\alpha}(x,t)~\delta(v-u_{\alpha}(x,t))~, \label{multistream}
\ee
where $\delta$ stands for the Dirac delta. The streams
represent infinitely thin filaments in phase space. If $f$ obeys the
Vlasov equation, then the functions $n_{\alpha}$ and $u_{\alpha}$
each satisfy the following continuity and momentum conservation
equations:
\begin{eqnarray}
\label{continuity} \frac{\partial\,n_{\alpha}}{\partial\,t} +
\frac{\partial}{\partial\,x}(n_{\alpha}u_{\alpha}) &=& 0 \,, \label{multi1}\\
\label{momentum} \frac{\partial\,u_{\alpha}}{\partial\,t} +
u_{\alpha}\frac{\partial\,u_{\alpha}}{\partial x} &=&
\frac{e}{m}\frac{\partial\phi}{\partial\,x}~, \label{multi2}
\end{eqnarray}
coupled of course to Poisson's equation with the electron density
given by $n(x,t) = \sum_\alpha p_\alpha n_\alpha(x,t)$. Note that
the representation (\ref{multistream}) presents some drawbacks, as
the functions $u_{\alpha}(x,t)$ can become multivalued during the
time evolution. This means that the system
(\ref{multi1})--(\ref{multi2}) will develop singularities, such as
an infinite density at certain positions. When this happens, the
fluid description (\ref{multi1})--(\ref{multi2}) ceases to be
valid, although the phase space picture of the streams is still
correct.

This line of reasoning can be extended to the quantum case
\cite{Haas,Anderson} by applying the Madelung representation of
the wave function to the system (\ref{sp1})--(\ref{sp2}).  Let us
introduce the real amplitude $A_{\alpha}(x,t)$ and the real phase
$S_{\alpha}(x,t)$ associated to the pure state $\psi_{\alpha}$
according to
\begin{equation}
\label{as} \psi_{\alpha} = A_{\alpha}\exp({i\,S_{\alpha}/\hbar})
\,.
\end{equation}
The density $n_{\alpha}$ and the velocity $u_{\alpha}$ of each
stream are given by
\begin{equation}
\label{nu} n_{\alpha} =|\psi_{\alpha}|^2= A_{\alpha}^2 \,,\qquad
u_{\alpha} = \frac{1}{m}\frac{\partial\,S_{\alpha}}{\partial\,x}
\,.
\end{equation}
Introducing Eqs. (\ref{as})--(\ref{nu}) into Eqs.
(\ref{sp1})--(\ref{sp2}) and separating the real and imaginary
parts of the equations, we find
\begin{eqnarray}
\label{continuity2} \frac{\partial\,n_{\alpha}}{\partial\,t} +
\frac{\partial}{\partial\,x}(n_{\alpha}u_{\alpha}) &=& 0 \,,\\
\label{momentum2} \frac{\partial\,u_{\alpha}}{\partial\,t} +
u_{\alpha}\frac{\partial\,u_{\alpha}}{\partial x} &=&
\frac{e}{m}\frac{\partial\phi}{\partial\,x} +
\frac{\hbar^2}{2m^2}\frac{\partial}{\partial\,x}
\left(\frac{\partial^{2}(\sqrt{n_{\alpha}})/\partial\,x^2}{\sqrt{n_{\alpha}}}\right)
\,.
\end{eqnarray}
Quantum effects are contained in the $\hbar$-dependent term in
(\ref{momentum2}) (sometimes called the Bohm potential). If we set
$\hbar = 0$, we obviously obtain the classical multistream model
(\ref{multi1})--(\ref{multi2}). An attractive feature of the
quantum multistream model is that, contrarily to its classical
counterpart, it does not generally develop singularities. This is
thanks to the Bohm potential, which, by introducing a certain
amount of wave dispersion, prevents the density to build up
indefinitely.

Linearizing equations (\ref{continuity2})--(\ref{momentum2})
(supplemented by Poisson's equation) around the spatially
homogeneous equilibrium: $n_{\alpha} = n_0$, $u_{\alpha} =
u_{0\alpha}$ and $\phi=0$, one obtains the following dielectric
constant
\begin{equation}
\label{dispmulti} \varepsilon_{\rm H}(\omega, k) = 1 -
\sum_{\alpha=1}^{N} p_{\alpha} ~\frac{\omega_{p}^2}{(\omega - k
u_{0\alpha})^{2} - \hbar^{2}k^{4}/4m^{2}}.
\end{equation}
The classical multistream relation is obtained simply by  setting
$\hbar=0$ in (\ref{dispmulti}).

The equivalence of the Wigner-Poisson and Hartree models can be
readily proven on the linear dispersion relation. The homogeneous
equilibrium described above corresponds to wave functions \be
\psi_{\alpha} = \sqrt{n_0}~\exp\left(i \frac{m
u_{0\alpha}}{\hbar}~x\right). \label{equipsi}\ee
The Wigner transform (\ref{wigfunc}) of the above wave functions
(\ref{equipsi}) is given by the following expression
\be f_0(v) = \sum_{\alpha=1}^{N} p_{\alpha} n_0
\delta(v-u_{0\alpha}), \label{f0} \ee
where $\delta$ stands for the Dirac delta. By inserting (\ref{f0})
into the WP dispersion relation (\ref{dispwp1}), one recovers
precisely the multistream dispersion relation (\ref{dispmulti}).

\subsection{Fluid model} \label{sec:fluid}
In classical plasma physics, fluid (or hydrodynamic) models are
often derived by taking moments of the appropriate kinetic
equation (e.g. Vlasov's equation) in velocity space. The moment of
order $s$ is defined as: \be M_s(x,t) = \int_{-\infty}^{+\infty}
f(x,v,t) v^s dv. \ee Then, the zeroth order moment is the spatial
density and obeys the continuity equation; the first order moment
is the average velocity and obeys a momentum conservation
equation; the second order moment is related to the pressure, and
so on. This procedure generates an infinite number of fluid
equations, which is usually truncated at a relatively low order by
assuming an appropriate closure equation. The closure often takes
the form of a thermodynamic equation of state, relating the
pressure to the density, e.g. the polytropic relation $P \propto
n^\gamma$. The same procedure can be applied to the Wigner
equation, although some steps in the derivation are somewhat
subtler than in the classical case. With this technique, quantum
fluid equations were derived in \cite{qfluid}. Here, we shall
present a succinct derivation of the same fluid model using a
different approach based on the Hartree equations \cite{Gasser,
Gardner}.

The starting point is the system of $2N$ equations
(\ref{continuity2})--(\ref{momentum2}), which is completely
equivalent to Hartree's model (\ref{sp1})--(\ref{sp2}).
Let us define the global density $n(x,t)$
\be
n(x,t) = \sum_{\alpha=1}^{N} p_\alpha n_\alpha
\ee
and the global average velocity $u(x,t)$
\be
u(x,t) \equiv \langle u_\alpha \rangle = \sum_{\alpha=1}^{N}
p_\alpha \frac{n_\alpha}{n}~u_\alpha
\ee
By multiplying the continuity equation (\ref{continuity2}) by
$p_\alpha$ and summing over $\alpha=1\dots N$, we obtain \be
\label{cont3} \frac{\partial\,n}{\partial\,t} + \frac{\partial (n
u)}{\partial x} = 0 \ee
Similarly, for the equation of momentum conservation
(\ref{momentum2}), one obtains \be \label{mom3}
\frac{\partial\,u}{\partial\,t} + u \frac{\partial u}{\partial x}
= \frac{e}{m}\frac{\partial \phi}{\partial x} +
\frac{\hbar^2}{2m^2} ~\frac{\partial}{\partial x}
\sum_{\alpha=1}^{N} p_\alpha \left( \frac{\partial_x^{2}
\sqrt{n_\alpha}}{\sqrt{n_\alpha}} \right) -\frac{1}{mn}
\frac{\partial P}{\partial x} \ee
where the pressure $P(x,t)$ is defined as
\be
P = mn \left[\frac{\sum_{\alpha} p_\alpha n_\alpha u^2_\alpha}{n}
- \left(\frac{\sum_\alpha p_\alpha n_\alpha u_\alpha}{n}\right)^2
\right] \equiv mn (\langle u_\alpha^2 \rangle - \langle u_\alpha
\rangle^2)
\ee

So far the derivation is exact, but (\ref{mom3}) still involves a
sum over the $N$ states, so no simplification was achieved. Our
purpose is to obtain a closed system of two equations for the {\it
global} averaged quantities $n$ and $u$. In order to close the
system, two approximations are needed:
\begin{enumerate}
\item
We postulate a classical equation of state, relating the pressure
to the density: $P = P(n)$.
\item
We assume that the following substitution is allowed: \be
\sum_{\alpha=1}^{N} p_\alpha \left( \frac{\partial_x^{2}
\sqrt{n_\alpha}}{\sqrt{n_\alpha}} \right) ~\Longrightarrow~
\frac{\partial_x^{2} \sqrt{n}}{\sqrt{n}} \ee
\end{enumerate}
It can be shown that the second hypothesis is satisfied for length
scales larger than $\lambda_F$. This will be apparent from the
linear theory detailed in Sec. \ref{sec:linear}.


With these assumptions, we obtain the following reduced system of
fluid equations for the global quantities $n$ and $u$
\be \label{cont4} \frac{\partial\,n}{\partial\,t} + \frac{\partial
(n u)}{\partial x} = 0 \ee
\be \label{mom4} \frac{\partial\,u}{\partial\,t} + u
\frac{\partial u}{\partial x} = \frac{e}{m}\frac{\partial
\phi}{\partial x} + \frac{\hbar^2}{2m^2} ~\frac{\partial}{\partial
x} \left( \frac{\partial_x^2 \sqrt{n}}{\sqrt{n}} \right)
-\frac{1}{mn} \frac{\partial P}{\partial x} \ee
where $\phi$ is given by Poisson's equation. We stress that we
have transformed a system of $2N$ equations
(\ref{continuity2})--(\ref{momentum2}) into a system of just two
equations.

An interesting form of the system (\ref{cont4})--(\ref{mom4}) can
be obtained by introducing the following `effective' wave function
\begin{equation}
\Psi(x,t) = \sqrt{n(x,t)}\exp{(iS(x,t)/\hbar)},
\end{equation}
where $S(x,t)$ is defined according to the relation $mu =
\partial_x S$, and $n=|\Psi|^2$. It is easy to show that
(\ref{cont4})--(\ref{mom4}) is equivalent to the following
nonlinear Schr\"odinger equation
\begin{equation}
\label{nlse} i\hbar\frac{\partial\Psi}{\partial\,t} = -
\,\frac{\hbar^2}{2m}~\frac{\partial^2 \Psi}{\partial x^2} -
e\phi\Psi + W_{{\rm eff}}(|\Psi|^2)~ \Psi.
\end{equation}
$W_{{\rm eff}}(n)$ is an effective potential related to the
pressure $P(n)$
\begin{equation}
\label{weff} W_{\rm eff}(n) =
\int^{n}\frac{dn'}{n'}\frac{dP(n')}{dn'}.
\end{equation}

As an example, let us consider a one-dimensional (1D)
zero-temperature fermion gas, whose equation of state is a
polytropic with exponent $\gamma = 3$
\begin{equation}
\label{pdeg} P = \frac{mv_{F}^2}{3n_{0}^2}n^3,
\end{equation}
where $v_F$ is the Fermi velocity computed with the equilibrium
density $n_0$. Using (\ref{pdeg}), the effective potential becomes
\begin{equation}
\label{weff2} W_{\rm eff} = \frac{mv_{F}^{2}}{2n_{0}^{2}}|\Psi|^4
\,.
\end{equation}

This fluid model is a useful approximation, as it reduces
dramatically the complexity of the Hartree system ($2N$ equations)
or the Wigner equation (phase space dynamics). Its validity is
limited to systems that are large compared to $\lambda_F$. Like
all fluid approximations, it neglects typical kinetic phenomena
originating from the details of the phase space distribution
function. In particular, it cannot reproduce Landau's
collisionless damping.

\subsection{Vlasov-Poisson}
\label{sec:Vlasov} The Wigner and Hartree approaches are both {\it
kinetic} and {\it quantum}. The above fluid model was derived by
dropping kinetic effects, while preserving quantum effects through
the Bohm potential. Another possible approximation could consist
in neglecting quantum effects while keeping kinetic ones. The
resulting semiclassical limit is, of course, given by the Vlasov
equation
\be \frac{\partial{f}}{\partial{t}} +
v\frac{\partial{f}}{\partial{x}}- \frac{e}{m} \frac{\partial
\phi}{\partial{x}} \frac{\partial{f}}{\partial{v}} = 0,
\label{vlasov} \ee coupled, as usual, to Poisson's equation.

The Vlasov-Poisson (VP) system has been used to study the dynamics
of electrons in metal clusters and thin metal films
\cite{Calvayrac,Manf-Herv}. It is appropriate for large excitation
energies, for which the electrons' de Broglie wavelength is
relatively small, thus reducing the importance of quantum effects
in the electron dynamics. However, as we have seen in Sec.
\ref{sec:quantum plasmas}, the equilibrium distribution for
electrons in metals lies deeply in the quantum region ($g_Q
\gtrsim 1$), so that the initial condition must be given by a
quantum Fermi-Dirac distribution. In this sense, the VP system is
{\it semiclassical}.

\subsection{Initial and boundary conditions}
In order to perform numerical simulations, boundary and initial
conditions must be specified. Physically, the initial condition
should represent a situation of thermodynamics equilibrium (ground
state). For electrons in metals, most analytical results are
obtained in the case of an infinite system (often referred to as
the `bulk' in solid state physics), which can be realized in
practice by taking periodic boundary conditions with spatial
period $L$ (this, of course, introduces a lower bound for the wave
numbers, namely $k_0=2\pi/L$). For such an infinite system, the
ground state can be easily specified. For the WP and VP models,
any function of the velocity only $f_0(v)$ constitutes a
stationary state. In particular, for fermions we take the
Fermi-Dirac distribution (see, however, Sec. \ref{sec:FD3D} for a
discussion on the appropriate Fermi-Dirac distribution for 1D
problems):
\be f_0(v)= {\rm const.} \times \frac{1}{1+e^{\beta_e
(\epsilon-\mu)}}, \label{fd} \ee where $\beta_e=1/k_B T_e$ and
$\epsilon=mv^2/2$ is the single-particle energy (here, $v$ is the
modulus of the 3D velocity vector). The chemical potential
$\mu(T)$ is determined so that $\int f_0 dv =n_0$, where $n_0$ is
the equilibrium density; $\mu$ becomes equal to the Fermi energy
$E_F$ in the limit $T \to 0$.
The same ground state can be defined for the Hartree model by
choosing the initial wave functions in the form (\ref{equipsi}).
The Fermi-Dirac distribution is then specified by the
probabilities $p_\alpha = [1+e^{\beta_e
(\epsilon_\alpha-\mu)}]^{-1}$, with $\epsilon_\alpha = m
u_{0\alpha}^2/2$.

The bulk approximation is not relevant for metal clusters and
other metallic nanostructures, which are small isolated objects
that exist either in vacuum or embedded in a background
non-metallic matrix. For open quantum systems (Wigner or Hartree),
the choice of appropriate boundary conditions is a subtle issue,
which will not be addressed here \cite{Frensley}. For the
semiclassical VP system, one can, for instance, use open
boundaries for the Vlasov equation (zero incoming flux) and
Dirichlet conditions for the Poisson equation.

As to the initial condition, the ground state is easily determined
for the VP system with open boundaries. The main difference from
the bulk equilibrium is that, for an open system, the
electrostatic energy does not vanish and the equilibrium density
$n_0(x)$ is position dependent. As any function of the total
energy is a stationary solution of the Vlasov equation, we define
the initial state as a Fermi-Dirac distribution (\ref{fd}) with
$\epsilon(x,v)=mv^2/2-e\phi(x)$, where the potential $\phi$ is not
yet known. By plugging this Fermi-Dirac distribution into
Poisson's equation, one obtains a nonlinear equation that can be
solved iteratively to obtain $\phi$, which in turns yields the
ground state distribution.

In contrast, stationary solutions of the Wigner equations are not
simply given by functions of the energy, so that the above
procedure cannot be applied. It is easier to compute the ground
state in terms of the Hartree wave functions $\psi_\alpha$, and
then compute the corresponding Winger function with
(\ref{wigfunc}). Several methods to compute the ground state wave
functions in the Hartree formalism are available in the literature
\cite{Kohn,Ekardt,Parr} and will not be discussed here.

\section{Linear theory}\label{sec:linear}
In order to compare the various models described in Sec.
\ref{sec:Models}, we shall go into some details of the linear
theory for a zero-temperature homogeneous equilibrium with
periodic boundaries, both in 1D and in 3D. More detailed
calculations on the linear theory of quantum plasmas can be found
in \cite{Pines,balescu,Bonitz2}.

\subsection{Zero-temperature 1D Fermi-Dirac
equilibrium}\label{sec:FD1D} In one spatial dimension, the
Fermi-Dirac distribution at $T=0$ is given by $f_{0}(v) =
n_{0}/2v_{F}$ if $|v| \le v_{F}$ and $f_{0}(v) = 0$ if $|v| >
v_{F}$. The Fermi velocity in 1D is \be v_F = \frac{\pi}{2}
\frac{\hbar n_0}{m}. \ee This distribution is identical to the
so-called `water-bag' distribution, which has been extensively
used in classical plasma physics \cite{waterbag}.
Using the Wigner linear dielectric constant (\ref{dispwp2}) and
developing the results in powers of $k$ and $\hbar$, one obtains
the following dispersion relation (details are given in
\cite{qfluid})
\begin{equation}
\label{dispwp3} \omega^2 = \omega_p^{2} + k^{2}v_{F}^2 +
\frac{\hbar^2 k^4}{4 m^2} + \frac{\hbar^2 k^6 \lambda_{F}^2}{3
m^2} + O(\hbar^4, k^{12}) \,.
\end{equation}
With the Vlasov dielectric constant, the dispersion relation for
the same equilibrium reads as
\begin{equation}
\label{dispvp3} \omega^2 = \omega_p^{2} + k^{2}v_{F}^2  \,.
\end{equation}
Note that the above Vlasov dispersion relation is {\it exact}: no
terms of order $k^4$ or higher exist. We also point out that, for
this 1D equilibrium, the imaginary part of the dielectric constant
vanishes identically and therefore there is no Landau damping.

We now want to compare the above result (\ref{dispwp3}), obtained
with the `exact' Wigner-Poisson model, with the equivalent result
obtained with the fluid model developed in Sec. \ref{sec:fluid}.
Let us consider the fluid equations (\ref{cont4})-(\ref{mom4}) in
the case of a zero-temperature 1D fermion gas, for which the
pressure is given by (\ref{pdeg}). Linearizing around the
homogeneous equilibrium $n=n_0$, $u=\phi=0$, one obtains the
following dispersion relation (no further approximation has been
used)
\begin{equation}
\label{dispfluid} \omega^2 = \omega_p^{2} + k^{2}v_{F}^2 +
\frac{\hbar^2 k^4}{4 m^2} \,.
\end{equation}
We see that the fluid (\ref{dispfluid}) and the Wigner
(\ref{dispwp3}) dispersion relations coincide up to terms of order
$k^4$. This confirms our conjecture that the fluid model is a good
approximation of the WP (or Hartree) model in the limit of long
wavelengths.

\subsection{Zero-temperature 3D Fermi-Dirac
equilibrium}\label{sec:FD3D}
From a physical point of view, it is possible to imagine a real
physical system that displays 1D behavior. This can be realized,
for instance, in a thin metal film where the dimensions parallel
to the surfaces are much larger than the thickness of the film.
The electron dynamics can then be described by a 1D infinite slab
model depending only on the co-ordinate $x$ normal to the film
surfaces. Even in such a situation, however, the 1D Fermi-Dirac
distribution discussed in Sec. \ref{sec:FD1D} is not realistic.
Indeed, in the ground state at zero temperature, the electrons
occupy all the available quantum states up to the Fermi surface,
defined by $\vline~ v ~\vline \le v_F$. There is no reason why
states with $v_y \neq 0$ and $v_z \neq 0$ should not be available,
and indeed they are occupied. Therefore, the equilibrium
distribution {\it is always three-dimensional}, even in a 1D
infinite slab geometry: \be \label{fd3d} f_0^{\rm 3D}(v) =
\frac{n_0}{\frac{4}{3} \pi v_F^3}~{\rm for}~|v| \le v_F, ~~~~{\rm
and} ~~~~ f_0^{\rm 3D}(v) = 0~{\rm for}~ |v|> v_{F}. \ee
However, it is still possible to keep the 1D geometry for the
Vlasov and Poisson's equations, provided that one uses, as initial
condition, the 3D Fermi-Dirac distribution (\ref{fd3d}) projected
on the $v_x$ axis: $f_0^{\rm 1D}(v_x) = \int \int f_0^{\rm 3D}
dv_y dv_z$. This yields (we now write $v$ for $v_x$)
\begin{equation}
f_0^{\rm 1D}(v) = \frac{3}{4}\frac{n_0}{
v_F}\left(1-\frac{v^2}{v_F^2}\right)~{\rm for}~|v| \le v_F,
~~~~{\rm and} ~~~~ f_0^{\rm 1D}(v) = 0~{\rm for}~ |v|> v_{F}.
\label{fd3D1}
\end{equation}

This approach is not as contrived as it might appear at first
sight. Indeed, linear wave propagation in a collisionless plasma
is intrinsically a 1D phenomenon, which involves plane waves
traveling in a well-defined direction. In computing the dispersion
relation from the 3D equivalent of (\ref{dispwp2}) or
(\ref{dispvp}), we would first integrate over the two velocity
dimensions normal to the direction of wave propagation (which can,
arbitrarily and without loss of generality, be chosen to be the
$v_x$ direction). Therefore, we would be left with a reduced
distribution such as (\ref{fd3D1}) that intervenes in a 1D
dispersion relation such as (\ref{dispwp2}) or (\ref{dispvp}).
This line of reasoning is no more valid when nonlinear effects
become important, as these may trigger truly 3D phenomena.
Collisions also constitute an intrinsically 3D effect.

We now insert (\ref{fd3D1}) into (\ref{dispwp2}) or (\ref{dispvp})
and compute the dispersion relations for the WP and VP systems,
developed in powers of $k$. For the equilibrium (\ref{fd3D1}), the
dielectric constant does display an imaginary part, and therefore
there is, in principle, some collisionless damping associated with
this equilibrium; however, we shall neglect it for the moment and
concentrate on the real part of $\varepsilon(k,\omega)$. We also
assume the following ordering \be \frac{\hbar k}{m} \ll v_F \ll
\frac{\omega}{k}, \label{ordering} \ee
which is valid for long wavelengths. Note that the second
inequality in (\ref{ordering}) means that the phase velocity of
the wave must be greater than the Fermi velocity. With this
assumption, the dispersion relation up to fourth order in $k$
reads as
\be \label{dispwp4}\omega^2 = \omega_p^{2} + \frac{3}{5}~
k^{2}v_{F}^2 + \frac{\hbar^2 k^4}{4 m^2} + \dots \,. \ee

Let us now derive the dispersion relation for the quantum fluid
model (\ref{cont4})-(\ref{mom4}), by assuming an equation of state
of the form:
\be \frac{P}{P_0} = \left(\frac{n}{n_0}\right)^\gamma ,
\label{eqstate} \ee
where $P_0$ and $n_0$ are the equilibrium pressure and density,
respectively. We obtain:
\begin{equation}
\label{dispfluid2}\omega^2 = \omega_p^{2} + \gamma k^{2} v_0^2 +
\frac{\hbar^2 k^4}{4 m^2} ~,
\end{equation}
where $v_0^2 = P_0/(m n_0)$. We note that the 1D fluid dispersion
relation (\ref{dispfluid}) is recovered correctly when $\gamma=3$
and $P_0 = n_0~mv_F^2/3$ as can be deduced from (\ref{pdeg}). Now,
in 3D, the pressure of a quantum electron gas at thermal
equilibrium and zero temperature can be written as \cite{Landau}
\be P_0 = \frac{2}{5} n_0 E_F \label{pressure3D} \ee
(where $E_F$ is computed at equilibrium), so that $v_0^2 =
v_F^2/5$ and (\ref{dispfluid2}) becomes
\begin{equation}
\label{dispfluid3}\omega^2 = \omega_p^{2} + \frac{\gamma}{5} k^{2}
v_0^2 + \frac{\hbar^2 k^4}{4 m^2} ~.
\end{equation}
One may think that the correct exponent to use in the equation of
state (\ref{eqstate}) is $\gamma=(D+2)/D$, yielding 5/3 in three
dimensions. However, with this choice, the fluid dispersion
relation (\ref{dispfluid3}) would differ from the WP result
(\ref{dispwp4}). The correct result is obtained by taking
$\gamma=3$, just like in the 1D case. Why is it so? The reason, as
was already mentioned earlier, is that linear wave propagation is
essentially a 1D phenomenon, because it involves propagation along
a single direction, without any energy exchanges in the other two
directions. The details of the linear dynamics are essentially
determined by the equation of state, which must therefore feature
the 1D exponent. In contrast, the {\it equilibrium} is truly 3D
(because we have projected the 3D Fermi-Dirac distribution over
the $x$ direction): therefore, the equilibrium pressure must
indeed be given by its 3D expression (\ref{pressure3D}).

\subsection{Finite temperature solutions: Landau damping}
So far, we have completely neglected the fact, discovered by
Landau in 1940 \cite{Landau-damp}, that electrostatic waves can be
exponentially damped even in the absence of any collisions. The
rigorous theory of Landau damping can be found in most plasma
physics textbooks. Here, we shall only remind the reader that the
damping originates from the singularity appearing in the
dispersion relations (\ref{dispwp1}) and (\ref{dispvp}) at the
point $v=\omega/k$ in velocity space. This corresponds to
particles whose velocity is equal to the phase velocity of the
wave $\omega/k$ (resonant particles). Landau showed that the
correct way to perform the integral in the dispersion relation is
not simply to take the principal value (which only yields the real
part of the frequency), but to integrate in the complex $v$ plane,
following a contour that leaves the singularity always on the same
side. With this prescription, the dielectric function is found to
possess an imaginary part, which in turn gives rise to a damping
rate $\gamma_L$ for the wave. This argument, originally developed
for the VP system, still holds for the quantum WP case, although
of course the numerical value of the damping rate will depend on
$\hbar$.

\begin{figure}[tb]
\begin{center}
\includegraphics*[width=6cm]{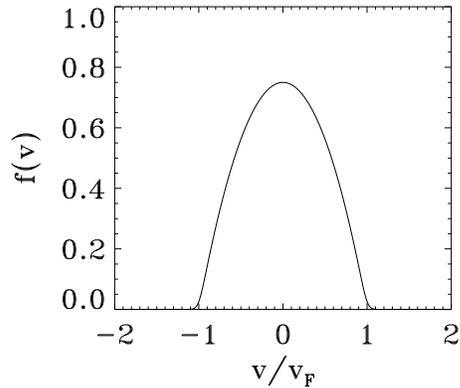}
\end{center}
\vspace{0pc} \caption{Fermi-Dirac velocity distribution projected
on a single velocity direction, with $T/T_F = 0.05$. }
\label{fig:fd3d}
\end{figure}

At zero temperature, no particles exist with velocity $v>v_F$.
Therefore, waves with phase velocity larger than $v_F$ are not
damped at all. For these waves, we have $k < \omega/v_F$; as the
real part of the frequency is approximately equal to the plasma
frequency, this means that waves for which $k \lambda_F < 1$ are
not damped. These are waves with a wavelength smaller than
$\lambda_F \equiv v_F/\omega_{pe}$, which is of the order of the
\AA ngstrom for metallic electrons (see table 1). At finite
temperature, the projected Fermi-Dirac distribution (\ref{fd3D1})
(see Fig. \ref{fig:fd3d}) extends to all velocities (although it
decays quickly), so that some amount of damping exists for all
wave numbers.

The linear damping rate can be computed from the dispersion
relation, see for instance \cite{Suh}, and we shall not deal with
this issue any further. It is more interesting to look for some
qualitative guess about the {\it nonlinear} phase that follows the
initial Landau damping \cite{Manfredi2}. Classically, Landau
damping lasts up to times of the order of the so-called bounce
time $\tau_b$, after which the evolution is inherently nonlinear.
The bounce time is related to the amplitude of the initial
perturbation of the equilibrium distribution
\be \label{pert} f(x,v,t=0^{+}) = f_0(v)~(1+\alpha \cos~kx), \ee
where $\alpha = \tilde{n}/n_0$ is the normalized density
perturbation and $k$ is the wave number of the perturbation. The
bounce time can then be written as: $\omega_p \tau_b =
\alpha^{-1/2}$. When the perturbation amplitude is not too small,
one generally observes that Landau damping stops after a time of
the order $\tau_b$. This happens because resonant particles (whose
velocity is close to the phase velocity of the wave) get trapped
inside the travelling wave, thus creating self-sustaining vortices
in the phase space. The presence of such vortices maintains the
electric field to a finite (albeit generally small) level.

\begin{figure}[tb]
\begin{center}
\includegraphics*[width=6cm]{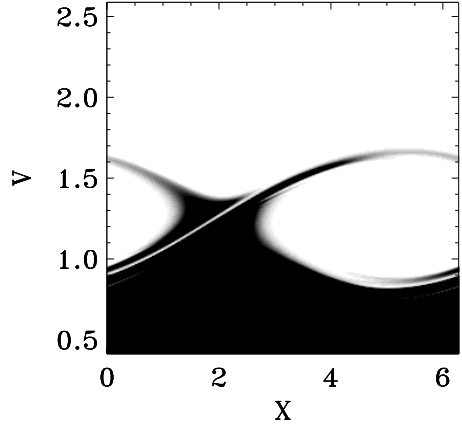}
\end{center}
\vspace{0pc} \caption{Classical phase space portrait of the
electron distribution function in the region around the phase
velocity of the wave. Position is normalized to $\lambda_F$ and
velocity to $v_F$. The simulation parameters are: $T = 0.01 T_F$,
$\alpha = 0.1$, and $k\lambda_F=1$. } \label{fig:phaseclass}
\end{figure}

We have performed numerical simulations of the VP system using a
Vlasov Eulerian code \cite{Cheng, Filbet}. The initial equilibrium
is given by the projected Fermi-Dirac distribution at finite
temperature
\begin{equation}
f_{e0}(x,v_x) = \frac{3}{4}\frac{n_0}{ v_F}\frac{T}{T_F}\ln
\left[1 + \exp \left(-\frac{\frac{1}{2}mv^2-\mu}{k_B T} \right)
\right], \label{fdwarm}
\end{equation}
which generalizes the zero-temperature result (\ref{fd3D1}). We
took an equilibrium temperature $T = 0.01 ~T_F$, whereas the
perturbation (\ref{pert}) is characterized by an amplitude $\alpha
= 0.1$ and a wave number $k\lambda_F=1$. The phase-space portraits
of the electron distribution (Fig. \ref{fig:phaseclass}) show, as
expected, the formation of a vortex traveling with a velocity
close to $\omega/k$. Further, it can be easily proven that the
extension of the vortex in velocity space $\delta v$ is related to
the perturbation's amplitude and wave number in the following way
\be \delta v = \frac{\omega_p}{k}~\alpha^{1/2}. \label{deltav}\ee

We now turn to the fully quantum case, described by the
Wigner-Poisson system with the same initial condition
(\ref{pert}). The simulations have been performed with the code
described in Ref. \cite{Suh}. As the wavelength of the
perturbation is $2\pi/k$, the classical phase-space vortex defines
a phase space area of order $\delta v /k$. If this area is smaller
than $\hbar/m$, then quantum mechanics forbids the creation and
persistence of such a structure \cite{Suh}. Using the relation
(\ref{deltav}), we find therefore that the phase-space vortex
should be suppressed by quantum effects when
\be \frac{\hbar k}{m} \gtrsim \frac{\omega_p}{k}~\alpha^{1/2}. \ee
Using normalized units, the above relation becomes
\be H ~k^2 \lambda_F^2 \gtrsim \sqrt{\alpha} , \label{vortex}\ee
where $H = \hbar \omega_p/E_F$ is a measure of the magnitude of
quantum effects. As quantum effects prevent particles from being
trapped inside the wave, we expect Landau damping to continue even
for times larger than the bounce time.

Physically, the above effect is related to quantum tunnelling:
particles trapped in the wave have a certain probability to be
de-trapped, even if their energy is less than that of the
potential well of the wave. Yet another way to picture this effect
is to consider that, if the potential well is too shallow (i.e.
for small $\alpha$), no quantum bound states can exist inside it.

\begin{figure}[tb]
\begin{center}
\includegraphics*[width=10cm]{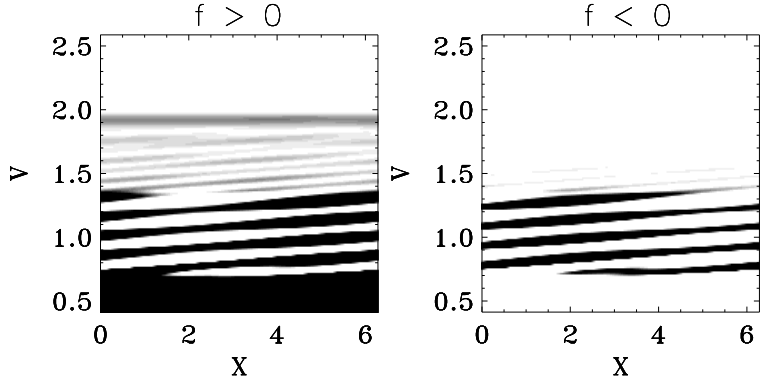}
\end{center}
\vspace{0pc} \caption{Quantum phase space portrait of the electron
Wigner function in the region around the phase velocity of the
wave. Position is normalized to $\lambda_F$ and velocity to $v_F$.
Same parameters as in Fig. \ref{fig:phaseclass}, with in addition
$H=1$. Left frame: positive part of $f(x,v)$; right frame:
negative part of $f$.} \label{fig:phasequan}
\end{figure}

\begin{figure}[tb]
\begin{center}
\includegraphics*[width=12cm]{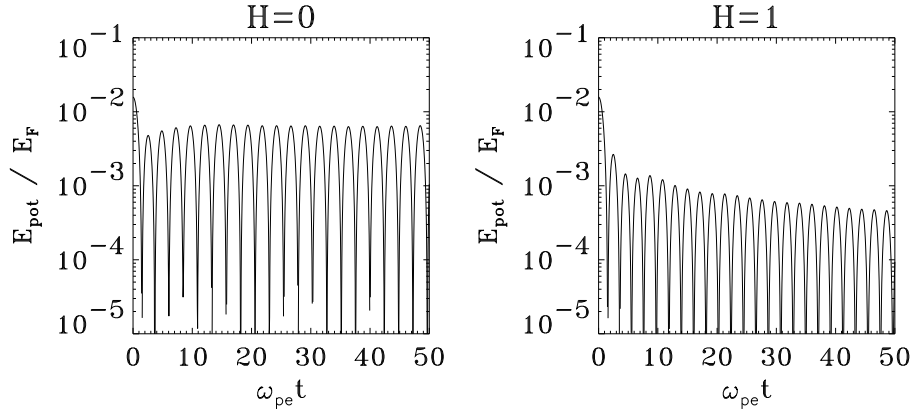}
\end{center}
\vspace{0pc} \caption{Time evolution of the potential energy,
normalized to the Fermi energy, for the classical case (left
frame) and the quantum case with $H=1$ (right frame).}
\label{fig:evolut}
\end{figure}

We have tested the above order-of-magnitude arguments by running
the same simulation as that of Fig. \ref{fig:phaseclass}, but
using the Wigner, instead of Vlasov, equation. We take $H=1$, so
that the inequality (\ref{vortex}) is satisfied. The phase space
portraits (Fig. \ref{fig:phasequan}) indeed confirm that no vortex
structures appear. Note also the appearance of large areas of
phase space where the Wigner distribution function is negative.
Comparing the classical and quantum evolution of the potential
energy (Fig. \ref{fig:evolut}), we observe that, for long times,
the electric field is significantly smaller in the quantum
evolution. These results  suggest that semiclassical models of the
dynamics of metallic electrons may underestimate the importance of
Landau damping.

\section{Conclusions and future developments}
In this paper, we have reviewed a number of kinetic and fluid
models that are appropriate to study the behavior of collisionless
plasmas when quantum effects are not negligible. The main
application of these models concerns the dynamics of electrons in
ordinary metals as well as metallic nanostructures (clusters,
nanoparticles, thin films). In particular, it is now possible to
study experimentally the electron dynamics on ultrafast
(femtosecond) time scales that correspond to the typical
collective plasma effects in the electron gas. The properties of a
quantum electron plasma (neutralized by the background ions) can
thus be measured with good accuracy and compared to the
theoretical predictions.

The main drawback of the models described in this paper is that
they all neglect electron-electron collisions. Strictly speaking,
the collisionless approximation should be valid only when $g_Q \ll
1$, which is not true for electrons in metals (see Table 1). As we
discussed in Sec. \ref{Pauli}, Pauli blocking does reduce the
effect of collisions for distributions near equilibrium, but many
interesting phenomena involve nonequilibrium electrons, so that
the validity of the collisionless approximation is not completely
clear. It is possible, in principle, to include collisional
effects in semiclassical models such as the Vlasov-Poisson system,
by simply adding a collision operator on the right-hand side of
the Vlasov equation \cite{Domps2}. The collision operator relevant
for a degenerate fermion gas is known as the \"{U}hling-Uhlenbeck
collision term and is basically a Boltzmann collision operator
that takes into account the exclusion principle. Even in the
semiclassical case, however, the validity of the
\"{U}hling-Uhlenbeck approach may be questioned for strongly
coupled plasmas ($g_Q \gtrsim 1$), for which it is conceptually
difficult to separate the mean-field from the collisional effects.

It is much harder to include a simple collisional term in truly
quantum models, such as the Wigner or Hartree equations
\cite{Lopez}. There is a vast literature on dissipative quantum
mechanics, but this is concerned with the interaction of a single
quantum particle with an external environment
\cite{Caldeira,Diosi,Zurek}. In addition, the coupling to the
environment is generally assumed to be weak, which is not the case
for metallic electrons. Recent approaches to the dynamics of
strongly coupled quantum plasmas range from quantum Monte Carlo
methods (for the equilibrium) to quantum molecular dynamics
simulations (for the nonequilibrium dynamics)
\cite{Bonitz,Ebeling}.

We shall mention another two possible extensions of the models
presented in this paper, but these are more technical points,
rather than conceptual issues such as the above-mentioned
inclusion of electron-electron collisions in the strongly coupled
regime.

The first issue concerns the coupling to the ion lattice. So far,
we have assumed that the ions form a motionless positively-charged
background described by the equilibrium density $n_0(x)$, which
may be position-dependent for open systems such as clusters. This
is appropriate for times shorter than the typical ion-electron
collision time (see Table 1), but for longer times the ion
dynamics must be taken into account. The ion dynamics, just like
the electrons', may be split into a mean-field part and a
collisional part. If we want to include only the mean-field
component (and neglect ion-electron collisions), then all we need
to do is add a Vlasov equation for the ionic species, supplemented
by a Maxwell-Boltzmann initial condition (as the ions are always
classical) -- see, for instance, \cite{Manf-Herv}. This is
conceptually simple, though it may require rather long simulation
times, because the typical ionic time scale (proportional to
$\omega_{pi}$) is much longer than the electrons'. As a first
approximation, electron-ion collisions may be modelled by a simple
relaxation term of the Bathnagar-Gross-Krook type \cite{BGK},
which is the kinetic analog of the Drude relaxation model of solid
state physics \cite{Ashcroft}. For a more accurate approach that
includes the full ion dynamics (mean-field and collisional), it
will be necessary to treat the ions with molecular dynamics
simulations.

Finally, one should consider the effect of magnetic fields (both
external and self-consistent) on the plasma dynamics. Magnetic
fields should not alter the main conclusions drawn in the present
work and are easily included in all the equations presented here.
For very strong fields, fusion and space plasma physicists have
developed a battery of approximations (guiding-center,
gyrokinetic, \dots) that allow one to reduce the complexity of the
relevant models. It is a challenging task to transpose these
methods to the physics of quantum plasmas \cite{Shokri,Shokri2}.
Magnetic fields should also trigger {\it spin} effects, which are
uniquely quantum-mechanical. The interaction of spin and Coulomb
effects, still a largely unexplored field, is thus likely to
become an active area of future research.

\medskip
\noindent {\bf Acknowledgments}\\ I would like to thank
Paul-Antoine Hervieux for his careful reading of the manuscript
and useful suggestions.


\end{document}